\documentclass[default]{aastex701}

\newcommand{\vdag}{(v)^\dagger}
\newcommand\aastex{AAS\TeX}
\newcommand\latex{La\TeX}
\shorttitle{Correlations between turbulent structures and dissipation in the magnetosheath}
\shortauthors{Doan et al.}
\usepackage{graphicx}
\usepackage{txfonts}

\usepackage[dvipsnames]{xcolor}
\usepackage{xcolor}

\usepackage{soul}

\newcommand{\lsv}[1]{{\bf \textcolor{purple}{#1$^{LSV}$}}}

\newcommand{\viet}[1]{{\bf \textcolor{red}{#1$^{V}$}}}

\newcommand{\vic}[1]{{\bf \textcolor{blue}{#1$^{VMC}$}}}

\newcommand{\aset}[1]{{\bf \textcolor{blue}{#1$^{AS}$}}}

\usepackage{hyperref}
\begin{document}

\title{Statistical correlations between turbulent structures and dissipation proxies in near-Earth space: \\the MMS 2023-2024 unbiased magnetosheath campaign}

\newcommand{\KTHaffil}{
Department of Electromagnetics and Plasma Physics, School of Electrical Engineering and Computer Science, KTH Royal Institute of Technology, Teknikringen 31, SE-11428 Stockholm, Sweden}

\author[orcid=0009-0004-1115-2410,sname='Doan']{Le Viet Anh Doan}
%\altaffiliation{These authors contributed equally to the article}
\affiliation{CPHT, CNRS, École polytechnique, Institut Polytechnique de Paris, 91120 Palaiseau, France}
\affiliation{\KTHaffil}
\email{le-viet-anh.doan@polytechnique.edu}

\author[orcid=0000-0002-5981-7758, sname='Sorriso-Valvo']{Luca Sorriso-Valvo} 
%\altaffiliation{These authors contributed equally to the article}
\affiliation{\KTHaffil}
\affiliation{Istituto per la Scienza e Tecnologia dei Plasmi, CNR, 70126 Bari, Italy}
\email[show]{lucsv@kth.se}

\author[orcid=0000-0000-0000-0001,sname='Raynal']{Pierre Raynal}
%\altaffiliation{These authors contributed equally to the article}
\affiliation{\KTHaffil}
\affiliation{Laboratoire de Physique des Plasmas (LPP), CNRS, Sorbonne Université, Observatoire de Paris, Université Paris-Saclay,
Ecole polytechnique, Institut Polytechnique de Paris, 91120 Palaiseau, France
}
\affiliation{Institut Supérieur de l’aéronautique et de l’espace, ISAE-Supaéro, Toulouse, France}
\email{fakeemail5@google.com}

\author[orcid=0000-0002-7848-9200, sname='Montagud-Camps']{Victor Montagud-Camps} 
\affiliation{Departament de Física, Universitat de les Illes Balears (UIB), 07122 Palma, Spain}
\affiliation{Institute of Applied Computing \& Community Code (IAC3), UIB, Spain}
\email{victor.montagud@uib.cat}

\author[orcid=0000-0003-1821-7390, sname=Settino]{Adriana Settino}
\affiliation{Space Research Institute, Austrian Academy of Sciences, 8042 Graz, Austria}
\email{Adriana.Settino@oeaw.ac.at}

\author[orcid=0000-0001-7633-2739, sname=Tamburrini]{Abiam Tamburrini}
\affiliation{Dipartimento di Fisica, Università della Calabria, 87036 Rende (CS), Italy}
\email{abiamtamburrini@gmail.com}

\author[orcid=0000-0003-1469-1116, sname='Svenningsson']{Ida Svenningsson} 
\affiliation{Chalmers University of Technology, Gothenburg, Sweden}
\email{ida.svenningsson@chalmers.se}

\author[orcid=0000-0001-8478-5797, sname='Chasapis']{Alexandros Chasapis} 
\affiliation{Laboratory for Atmospheric and Space Physics, University of Colorado, Boulder, CO, USA}
\email{Alexandros.Chasapis@lasp.colorado.edu}

\author[orcid=0000-0002-8335-1441, sname='Vo']{Tien Vo} 
\affiliation{Laboratory for Atmospheric and Space Physics, University of Colorado, Boulder, CO, USA}
\email{Tien.Vo@lasp.colorado.edu}

\author[orcid=0000-0002-9707-3147, sname='Yordanova']{Emiliya Yordanova} 
\affiliation{Swedish Institute of Space Physics, Uppsala, Sweden}
\email{emiliya.yordanova@irfu.se}

\author[orcid=0000-0003-3446-7322, sname='Richard']{Louis Richard} 
\affiliation{Rudolf Peierls Centre for Theoretical Physics, University of Oxford, Clarendon Laboratory, Parks Road, Oxford OX1 3PU, United Kingdom}
\email{louis.richard@physics.ox.ac.uk}

\author[orcid=0000-0003-4168-590X, sname='Pecora']{Francesco Pecora} 
\affiliation{Department of Physics and Astronomy, University of Delaware Newark, DE 19716, USA}
\email{fpecora@udel.edu}

\author[orcid=0000-0001-7597-238X, sname='V\"or\"os']{Zoltan V\"or\"os} 
\affiliation{Space Research Institute, Austrian Academy of Sciences, 8042 Graz, Austria}
\email{zoltan.voeroes@oeaw.ac.at}

\author[orcid=0000-0001-8184-2151, sname='Servidio']{Sergio Servidio} 
\affiliation{Dipartimento di Fisica, Università della Calabria, 87036 Rende (CS), Italy}
\email{sergio.servidio@fis.unical.it}

\author[orcid=0000-0003-3446-7322, sname='Pezzi']{Oreste Pezzi} 
\affiliation{Istituto per la Scienza e Tecnologia dei Plasmi, CNR, 70126 Bari, Italy}
\email{oreste.pezzi@istp.cnr.it}

%%%%\collaboration{all}{The Terra Mater collaboration}

\begin{abstract}

The terrestrial magnetosheath provides excellent opportunities for the experimental study of space plasma physics. 
Turbulence and kinetic-scale dissipation processes have been extensively studied separately. 
However, examining their cross-scale interplay may shed light on unsolved long-standing questions: which kinetic processes specifically dissipate the plasma turbulent fluctuations, and what drives them? 
In this study, we perform a statistical study of high-resolution data from the 2023–2024 \textit{unbiased campaign} of the Magnetospheric Multiscale (MMS) mission to examine correlations between local small-scale accumulation of turbulent energy and peaks of dissipation proxies. 
The results highlight that clear regional correlations between a vast ensemble of dissipation proxies emerge on a scale of a few ion inertial lengths. Furthermore, intermittent turbulent structures dissipate energy in a hierarchical ordering.

\end{abstract}



\section{Introduction} 
\label{sec:introduction}

%\lsv{NOTE: in the new-command definition section (at the beginning of the latex file) you can define your own text color for substantial modifications (please do not use it for typos, just fix it), just copy the same command as the present ones and change color name.}

\vskip 12pt

One of the challenges in the physics of space plasmas such as the solar wind and planetary magnetospheres is to explain the energy conversion and dissipation mechanisms responsible for plasma heating and particle energization \citep{2022ExA....54..427R}. Observations show that space plasmas remain significantly hotter than predicted from adiabatic expansion or collisional heating models \citep{Bruno2005, Marsch2006}. Since heliospheric plasmas are weakly collisional, turbulent fluctuations are thought to transfer energy from magnetohydrodynamic (MHD) scales toward kinetic scales, where it is then converted into internal energy by kinetic processes.% \citep{Verscharen20}. 
To investigate these processes, we focus on the terrestrial magnetosheath because it hosts strong and fully developed turbulence comparable to those of the solar wind.

Plasma fluctuations follow power-law spectra in the turbulent inertial range, then transition to kinetic turbulence with a spectral break near ion scales \citep{Bruno2005}. 
Below ion scales, the cascade continues toward electron scales, where dissipation becomes significant \citep{Sahraoui_2009, Alexandrova_2009}. This broadband cascade provides a pathway where energy conversion and, potentially, dissipation occur through a variety of mechanisms \citep{Matthaeus_2020,Pezzi_2021}, such as wave–particle interactions like Landau damping \citep{Howes_2011} and cyclotron resonance \citep{Marsch2006}, as well as nonlinear processes like stochastic heating \citep{Chandran_2010}. 
From both fluid and kinetic simulations, it is known that turbulent energy conversion is highly intermittent and tends to concentrate near coherent structures such as current sheets and reconnection sites \citep{Karimabadi_2013, Servidio_2015}. While these processes are well established in theory and simulations, their relative importance in observations still remains difficult to quantify.

Many studies have begun to link these processes to localized structures in turbulence. In the solar wind, current sheets are correlated with enhanced proton temperature and temperature anisotropy \citep{Osman_2012, Greco_2012}, and coherent structures have also been associated with enhanced suprathermal and energetic particle populations \citep{Tessein2013, Tessein2015}. Analysis tools for the identification of strong turbulent energy transfer also reveal that intermittent events are associated with localized and regional proton heating \citep{Sorriso_Valvo_2018, Yordanova_2021}. 
In the terrestrial magnetosheath, measurements reveal enhanced particle energization, and non-Maxwellian velocity distribution functions (VDF) near coherent structures \citep{Chasapis_2018, Sorriso_Valvo_2019, Bandyopadhyay_2020}. 
MHD simulations show that turbulence generates reconnecting current sheets that efficiently convert energy and accelerate particles \citep{Wan_2016, Franci_2017}, while kinetic simulations further show that electromagnetic work and pressure–strain interactions cluster around coherent structures \citep{Yang_2017}. 

Despite this progress, the causal relationship between turbulence and kinetic-scale energy conversion and dissipation remains unclear. Some studies interpret turbulence as the large-scale driver that forms strong gradients, current sheets, and reconnection sites where energy conversion is localized \citep{Karimabadi_2013, Matthaeus_2020}, while others emphasize more distributed wave–particle interactions \citep{Howes_2011, Howes_Klein_Li_2017}. Clarifying this relationship requires a statistical approach that directly connect turbulent energy transfer with dissipation signatures across varying plasma conditions.
Such approach has so far been limited. 
Many observational studies focus on intervals pre-selected for strong, visually identifiable events \citep{Osman_2012, Chasapis_2018}. Such targeted selections can enhance apparent correlations but are known to introduce selection biases \citep{Pecora_2023}.
Moreover, most analyses rely on a limited subset of diagnostics, even though different proxies capture distinct channels of energy conversion \citep{Chasapis_2017, Phillips_2022}. 
While recent works have begun combining multiple complementary diagnostics \citep{Yang_2018, Sorriso_Valvo_2019, Pezzi_2021}, systematic comparisons across several dissipation proxies under consistent observational conditions remain limited. 
Most importantly, these analyses did not account for the spatial and temporal displacement between turbulent structures and dissipation sites, resulting in weak correlations in direct point-to-point comparisons \citep{Yang_2018, Matthaeus_2020}. This behavior arises because dissipation effectively occurs in the vicinity of turbulent structures rather than exactly at their locations \citep{Greco_2012, Tessein2013, Sorriso_Valvo_2018}. 

In this study, we address these limitations by analyzing high-resolution data from the 2023–2024 Magnetospheric Multiscale (MMS) magnetosheath unbiased campaigns. 
Using this dataset, we systematically compare multiple complementary dissipation diagnostics and explicitly quantify their spatial displacement relative to turbulent energy-transfer events. 

\section{The 2023-2024 Magnetospheric MultiScale unbiased campaign} 

\begin{figure*}[ht!]
\includegraphics[width=\textwidth]{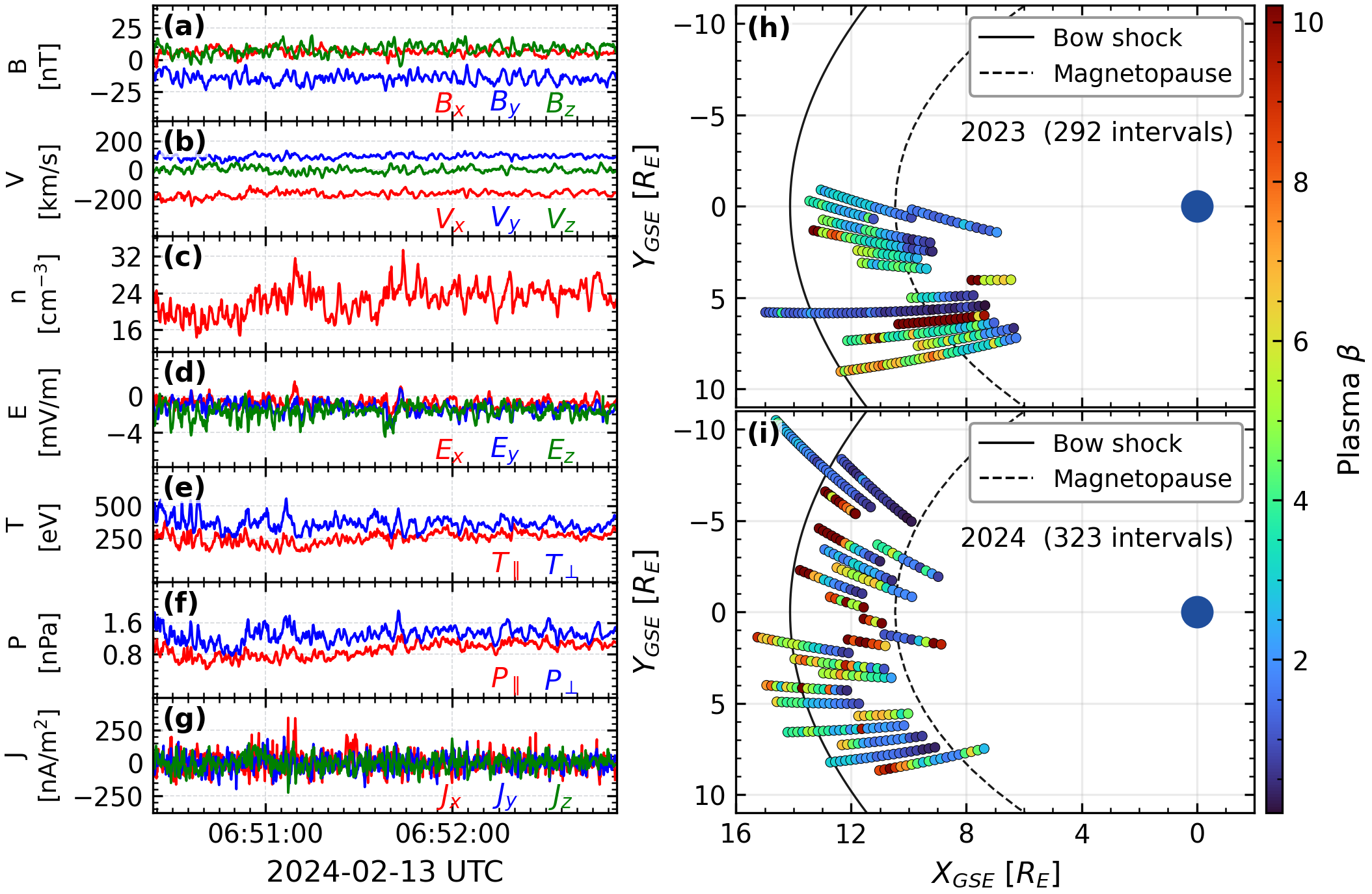}
\caption{Left: Example time series of plasma and field quantities parameters extracted from MMS1. Right: Position of each interval in the GSE plane, color-coded by the plasma $\beta$, for the 2023 (top) and 2024 (bottom) campaigns. The large blue circle represents the Earth. Magnetopause and bow shock are computed from \cite{farris_1994}, \cite{shue_1998}. }
\label{fig:dataset}
\end{figure*}

Burst-mode measurements from the four MMS spacecrafts are used to investigate turbulence and dissipation proxies in the terrestrial magnetosheath. The mean correlation length estimated from the magnetic-field autocorrelation was approximately $806\,\mathrm{km}$, significantly larger than the mean spacecraft separation of $33\,\mathrm{km}$. This implies that all four spacecrafts were sampling the same correlated plasma structure and supports the validity of the multi-spacecraft gradient estimates.

The magnetic field $\mathbf{B}$ is provided by the Fluxgate Magnetometer \citep{Russell2016} and sampled at 128 Hz, while the electric field $\mathbf{E}$ is provided by the Electric Double Probe at 8192 Hz \citep{Ergun2016}. 
Ion velocity distribution functions ($f_i$) and corresponding ion moments, such as ion number density ($n_i$), bulk ion velocity ($\mathbf{v}_i$), ion pressure tensor ($\mathbf{P}_i$) and ion temperature tensor ($\mathbf{T}_i$), are obtained from the Fast Plasma Investigation (FPI) at 150 ms \citep{Pollock2016}, with bulk velocities corrected using FPI spintone data. For electrons, only the number density ($n_e$) derived from the FPI is used.

In standard MMS operations, burst-mode intervals are selected through a Scientist-In-The-Loop process \citep{Fuselier2016}, which biases observations toward specific conditions. The MMS 2023–2024 unbiased campaigns was designed to provide systematically sampled intervals of magnetosheath turbulence without event preselection. The resulting dataset contains 615 intervals\footnote{Data are publicly available at the \href{https://lasp.colorado.edu/mms/sdc/public/}{Laboratory for Atmospheric and Space Physics}, Boulder, CO, USA} collected during 2 February–27 March 2023 \citep{Pecora2026ubmsh_list} and 13 March–27 April 2024. 
The campaign used repeating cycles of 2.5-3 min of burst data, followed by 7-9 min of fast data. 
These campaigns have supported statistical surveys of whistler waves \citep{Svenningsson_2024}, automated classification of plasma regimes \citep{Svenningsson_2025_ClassifyingMMS}, analyses of non-Maxwellian ion distributions \citep{Richard_2025_non-max}, and systematic studies of plasma and turbulence parameters \citep{pecora2025magnetosheath}. 
The left panels of Figure \ref{fig:dataset} show a representative interval from the 2024 campaign, characterized by steady conditions and strong turbulent fluctuations across all fields. The right panels summarize the spatial coverage of the 615 intervals. The broad range of plasma $\beta = 2\mu_0 n_i k_B T / |\mathbf{B}|^2 $ indicates that the dataset samples diverse turbulent environments throughout the magnetosheath. 

\begin{figure*}[t!]
\includegraphics[width=\textwidth]{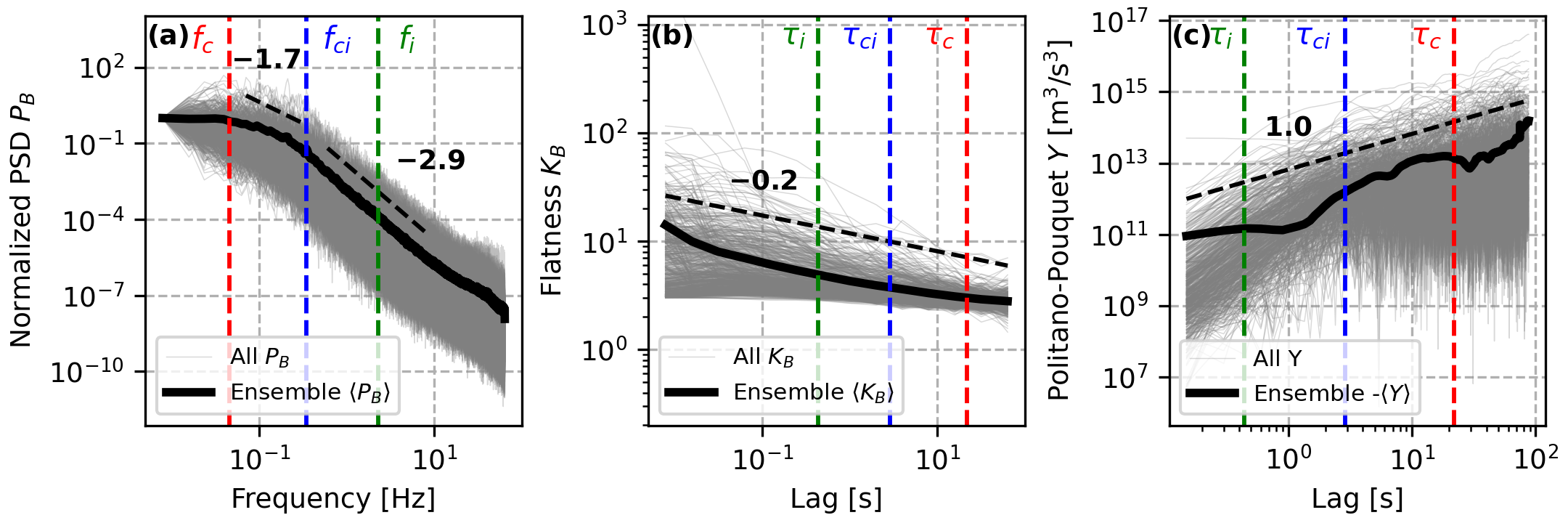}
\caption{Turbulence diagnostics from the unbiased campaigns. Each power spectral density is normalized to the power at the lowest frequency. Grey curves represent individual intervals, solid black curves for the ensemble averages, and dashed lines for the power-fits. The inertial range is bounded by $f_{\mathrm{c}}$ and $f_{ci}$, while the kinetic range extends beyond $f_i$, all indicated by colored vertical dashed lines. Converted time scales are respectively $\tau_{\mathrm{corr}}$, $\tau_{\rho i}$, and $\tau_i$ (same style).}
\label{fig:general}
\end{figure*}

To verify that the studied intervals contain fully developed turbulence, we compute standard diagnostics, summarized in Figure~\ref{fig:general}. 
The inertial range is bounded by the correlation frequency $f_{c}$ and the ion cyclotron frequency $f_{ci} = e|\mathbf{B}|/{2\pi m_i}$. The correlation frequency is obtained from the 1/e decay of the magnetic-field autocorrelation function \citep{Matthaeus_1999}. The kinetic range extends toward the ion inertial scale $d_i = c/\sqrt{n_i e^2/(\varepsilon_0 m_i)}$, with $m_i$ the ion mass. The computed root-mean-square velocity fluctuations satisfy $\sqrt{\left\langle \left| \mathbf{v} - \langle \mathbf{v} \rangle \right|^2 \right\rangle}/|\langle \mathbf{v} \rangle| < 0.2$ and supports the validity of the Taylor hypothesis. This scale is converted to the frequency $f_{i}$ through the mean plasma bulk flow velocity of each interval under the Taylor hypothesis \citep{Taylor_1938}.
The trace magnetic power spectral density (PSD) $P_B(f)$ is computed for each interval using the Welch method \citep{Welch_1967}, and then ensemble-averaged over all intervals, after normalization to the lowest mode power. 
Panel (a) shows that the PSD exhibits a persistent but very limited inertial range followed by a kinetic-scale spectrum, with exponents consistent with typical magnetosheath turbulence \citep{Yordanova2008,pecora2025magnetosheath}.

Intermittency is quantified using the scaling of the magnetic-field flatness $K_B(\tau)=\left\langle |\delta \mathbf{B}(\tau)|^{4}\right\rangle_t/\left\langle |\delta \mathbf{B}(\tau)|^{2}\right\rangle_t^{2}$, where $\delta \mathbf{B}(\tau)=\mathbf{B}(t+\tau)-\mathbf{B}(t)$, $\tau$ is the time lag , $\langle\cdot\rangle_t$ denotes a time average within each interval, and the result is averaged over the three magnetic-field components for visualization purposes (the three components showing similar scaling). 
Panel (b) shows that the flatness increases toward small scales and follows a power law, indicating non-Gaussian fluctuations and developed intermittency. 
The observed exponent is comparable to hydrodynamic turbulence \citep{Anselmet_1984} and quasi-perpendicular magnetosheath \citep{Gurchumelia2022}, but generally smaller than typical solar-wind and more turbulent magnetosheath values \citep{SorrisoValvo_Helios_2023,Quijia2021}, although significant interval-to-interval variability is present.
Finally, the MHD turbulent energy cascade is estimated using the Politano–Pouquet (PP98) third-order law \citep{Politano, MARINO20231}, formulated in terms of the magnetic fluctuations in Alfvén units $\mathbf{b}=\mathbf{B}/\sqrt{\mu_0 n_i m_i}$.
The mixed third-order structure function is defined as $Y(\tau) = \left\langle 
\delta v_\parallel(\tau)
\left( |\delta \mathbf{v}(\tau)|^2 + |\delta \mathbf{b}(\tau)|^2 \right)
- 2\, \delta b_\parallel(\tau)\,
\delta \mathbf{v}(\tau) \cdot \delta \mathbf{b}(\tau)
\right\rangle_t$ where longitudinal increments $\delta v_\parallel(\tau)$ and $\delta b_\parallel(\tau)$ are taken along each interval mean flow direction. 
\begin{comment}
Under homogeneous, stationary, incompressible MHD conditions, the PP law predicts a linear inertial-range scaling, $Y(\tau)\simeq -(4/3)\,\epsilon\,V_{\mathrm{sw}}\,\tau$, with $\epsilon$ the mean turbulent energy cascade rate.
\end{comment}
Panel (c) shows that the ensemble-averaged third-order moment follows a clear linear scaling across the inertial-range scales \citep{Hadid2018, Quijia2021}. 

Overall, the unbiased campaigns systematically exhibit clear signatures of developed turbulence, with an identifiable inertial range, scale-dependent intermittency, and a scaling consistent with the PP law prediction. For a more in-depth study of turbulent scales \citep{pecora2025magnetosheath}, see also analyses of the PP law validity and energy transfer rates \citep{Montagud_Camps_2026}.

\section{Proxies for turbulent structures and kinetic dissipation} 

\begin{comment}
    
Direct measurements of plasma heating are not available at kinetic scales, therefore several observational proxies have been proposed to identify regions where strong kinetic activity are expected. 
Since the spacecraft separation was shorter than the typical scales studied here, all proxies shown in Figure \ref{fig:Proxies} are computed using averaged measurements from the four MMS spacecraft to reduce instrumental noise in the particle distributions.
\end{comment}

A convenient way to quantify the scale-dependent transfer of energy in turbulent plasmas is through the Local Energy Transfer (LET) proxy, which originates from the MHD PP98 law \citep{Politano}. 
By expressing this law in a point-wise form, we obtain an estimator of the instantaneous nonlinear energy flux at scale $\ell$. The LET can be decomposed into physically interpretable contributions 
\begin{equation}
\mathrm{LET} = \left[|\mathbf{\delta v}|^{2}\,\delta v_{\parallel} + |\mathbf{\delta b}|^{2}\,\delta v_{\parallel} - 2\,\mathbf{(\delta v} \cdot \mathbf{\delta b})\,\delta b_{\parallel}\right]/\ell.
\end{equation}
In this study, the LET is evaluated at the scale $\ell \simeq d_i$, which marks the transition between MHD turbulence and ion-kinetic physics.
Studies have shown that intense LET events are correlated with enhanced nonlinear activity, often observed near current sheets and associated with kinetic distortions and plasma heating \citep{Sorriso_Valvo_2018, Yang_2018,Sorriso_Valvo_2019}. On the other hand, dissipation proxies can be estimated using the measured moments or the whole ion VDF.

The energy conversion between fields and particles is quantified through the electromagnetic work \citep{Zenitani_2011},
\[
\mathbf{J} \cdot \mathbf{E}'
=
\mathbf{J} \cdot
\left(
\mathbf{E}
+ \mathbf{v} \times \mathbf{B}
\right)\]
where $\mathbf{E'}$ denote the electric field measured in the co-moving frame of ions. 
This proxy isolates non-ideal field–particle interactions and identifies localized energization in kinetic turbulence \citep{Wan_2016, Yang_2018}.

Complementary information on how this energy is redistributed is provided by the pressure–strain interaction \citep{Del_Sarto_2016},
\[
(\mathbf{P} \cdot \nabla)\cdot \mathbf{v}
= P \theta + \mathbf{\Pi} : D,
\]
where $\theta=\nabla\cdot \mathbf{v}$, $P$ is the isotropic pressure, $\mathbf{D}$ is the traceless strain tensor, and $\mathbf{\Pi}$ is the deviatoric pressure tensor. The $P\theta$ term represents compressive work, while the $\mathrm{\Pi}-\mathrm{D}$ term captures shear-driven energy conversion. Additional filtering is applied to suppress residual spintone (see Appendix~\ref{sec:spintone}).
These proxies describe the transfer of bulk flow energy into internal energy and provide a measure of irreversible heating processes \citep{Yang_2017,Pezzi_2019}. 

\begin{figure*}[t]
\centering
\includegraphics[width=\textwidth]{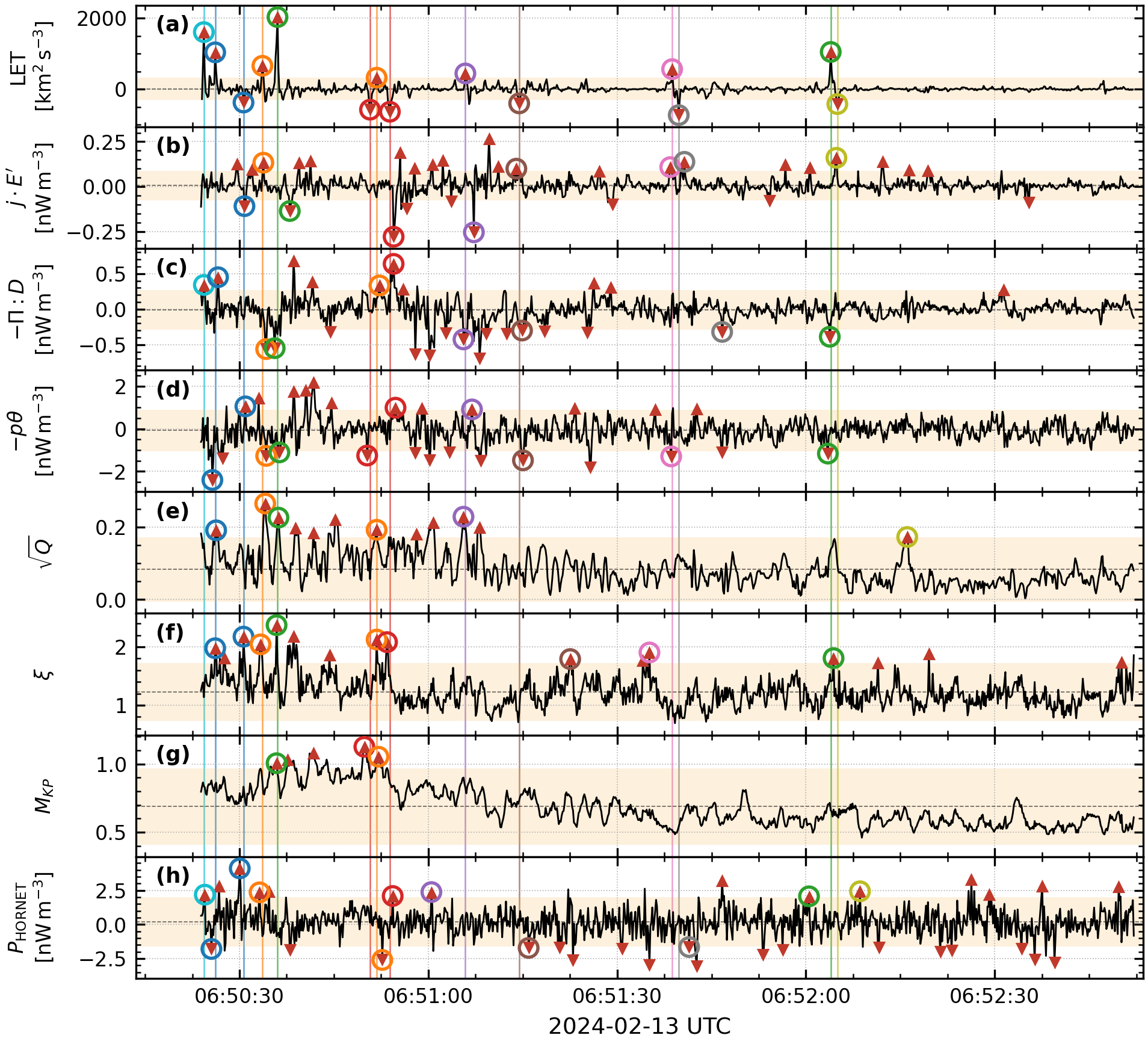}
\caption{Example time series of all computed proxies.  Detected events of turbulent or dissipation activity are marked with red stars. An example of event displacement matching analysis is performed between the LET and $\mathbf{j' \cdot E'}$, following the methodology in Appendix~\ref{sec:method_distance}. Identical green shapes mark a unique pair of matches, in which their amplitude and spatial distances are recorded for statistical analysis.}
\label{fig:Proxies}
\end{figure*}

Additional insights into the kinetic processes can also be obtained from the structure of the particle distribution. Departures from gyrotropy are quantified using the pressure agyrotropy parameter $\sqrt{Q}$ \citep{Swisdak_2016}, defined in a field-aligned coordinate with
$
\mathbf{P}=
\begin{pmatrix}
P_{\parallel} & P_{12} & P_{13} \\
P_{12} & P_{\perp} & P_{23} \\
P_{13} & P_{23} & P_{\perp}
\end{pmatrix}
$
as
$$
\sqrt{Q} = \sqrt{\frac{P_{12}^2 + P_{13}^2 + P_{23}^2}
{P_{\perp}^2 + 2P_{\parallel}P_{\perp}}} .
$$
A value $\sqrt{Q}=0$ corresponds to a gyrotropic distribution, while positive values indicate increasing departures from gyromotion around the magnetic-field direction and captures partial demagnetization \citep{Voros_2017,Che_2018}.

Departures from local thermal equilibrium (LTE) are quantified using the normalized non-Maxwellianity \citep{Greco_2012, Pezzi_2021}, which quantifies the deviation between the measured VDF $f_i$ and the corresponding Maxwellian $g_i$,
$$
\xi =
\frac{v_{th,i}^{3/2}}{n_i}
\sqrt{
\int_{\mathbb{R}^3} \bigl(f_i(\mathbf{v})-g_i(\mathbf{v})\bigr)^2\,\mathbf{d}^3\mathbf{v}},
$$
where $v_{th,i} = \sqrt{2k_BT/m_i}$ is the ion thermal speed. Larger values indicate possible velocity-space distortions, such as beams, heating \citep{Sorriso_Valvo_2019}, and other non-equilibrium kinetic features \citep{Graham}.

A complementary measure of departure from equilibrium is provided by the entropy deficit \citep{Kaufmann_2009, Liang_2019}. The kinetic entropy $s
= -k_B \int_{\mathbb{R}^3} f(\mathbf{v}) \ln f(\mathbf{v}) \,\mathbf{d}^3 \mathbf{v}
$ is compared with that of the corresponding Maxwellian distribution defined as $s_{M}
= (3/2)\,k_B n_i
\left[
1 + \ln\!\left(2\pi k_B T /m_i n_i^{2/3} \right)
\right]$, to define the normalized entropy deficit
$$
M_{KP}
=\frac{s_{M}-s}
{(3/2)\,k_B n_i}.
$$
Positive values indicate increasing departures from equilibrium with the presence of non-thermal velocity-space structure \citep{Argall_2022,Liang_2020}.

Finally, the HORNET proxy \citep{Barbhuiya_2024} characterizes the temporal evolution of velocity-space entropy through a Lagrangian derivative,
$$
P_{\mathrm{HORNET}}=-n_i T
\frac{\mathrm{d}}{\mathrm{d}t}
\left(\frac{s_{\mathrm{v},\mathrm{rel}}}{n_i}\right),
$$
with $
s_{ \mathrm{v},\mathrm{rel}}
=-k_B\!\int_{\mathbb{R}^3}\!f_i(\mathbf{v})
\ln\!\left({f_i(\mathbf{v})}/g_{i}(\mathbf{v})\right)
\,\mathrm{d}^3 \mathbf{v}$. 
Positive values indicate an ongoing evolution away from LTE, while negative values correspond to relaxation toward equilibrium. 

\begin{figure*}[t!]
\includegraphics[width=\textwidth]{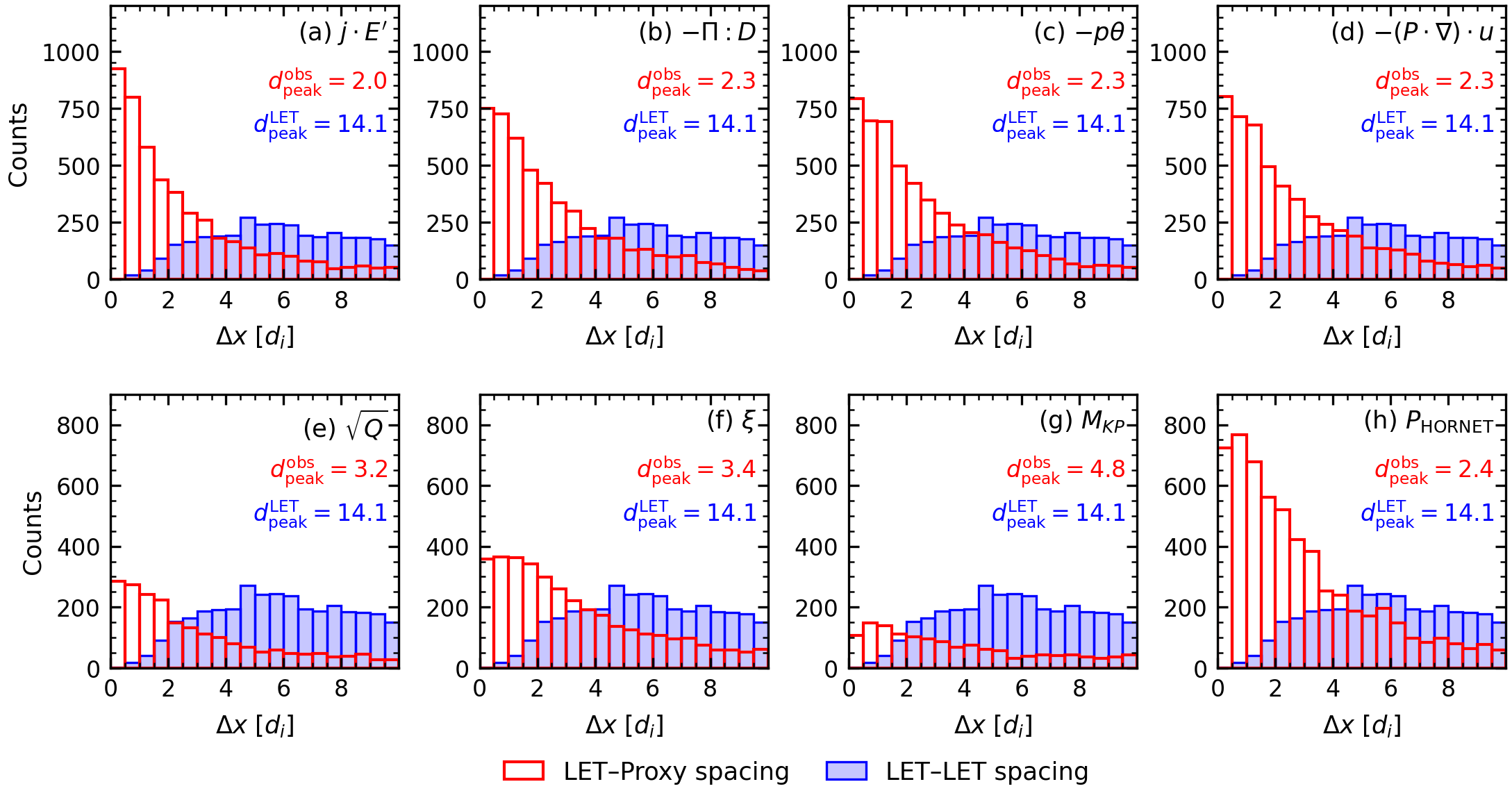}
\caption{Distributions of spatial separations between peaks of proxies. Red histograms show the separations between successive LET peaks with $d_{\mathrm{peaks}}^{\mathrm{LET}}$ the median of the distribution. Blue histograms show the separations between each LET peak and the nearest peak of the kinetic proxy, with $d_{\mathrm{peaks}}^{\mathrm{Proxy}}$ the median of the distribution.
All spatial distances are normalized by the ion inertial length $d_i$ of the corresponding interval.
\label{fig:co-location_pdf}}
\end{figure*} 

\section{Statistical relationship between turbulent structures and kinetic signatures} 
\label{sec:statistics}

To investigate whether the most intense kinetic dissipation occurs near intermittent turbulent structures, we initially quantify the spatial separation between each peak of the LET and the nearest peak of each dissipation proxy, using a peak-matching procedure (see Appendix \ref{sec:method_distance}). 
Figure~\ref{fig:co-location_pdf} shows the histograms of such LET-dissipation proxy distances, obtained from the whole unbiased campaign (red), compared to the distribution of distances between two successive peaks of the LET (blue). 
Note that all the dissipation proxies show a similar peak-to-peak distribution (not shown). 
In each panel, we also report the median of the distributions.

Figure~\ref{fig:co-location_pdf} shows that successive LET peaks are separated by a characteristic distance $d_{\mathrm{peak}}^{\mathrm{LET}}\simeq 11 d_i$ and indicates that intermittent turbulent structures are sparsely distributed at ion scales, typical of intermittent turbulence. 
In contrast, all kinetic proxies exhibit significantly smaller separations relative to LET peaks, suggesting that sites of dissipation might be correlated with the intermittent structures.
A systematic ordering is observed in the distributions. 
Energy-based proxies ($\mathbf{J}\cdot \mathbf{E'}$, $(\mathbf{P}\cdot\nabla)\cdot \mathbf{v}$, $P-\theta$, $\mathrm{\Pi}-\mathrm{D}$) generally show median separations of order $\sim 2 d_i$, while velocity-space proxies ($\sqrt{Q}$, $\xi$, $M_{KP}$) display broader distributions with median separations of order $\sim 3-4\,d_i$. 
This suggests a more local effect of turbulent structures on the former, and slightly more regional on the latter.
Interestingly, despite the HORNET $P_{\mathrm{HORNET}}$ being derived from the VDF, it displays characteristic separations comparable to those of the energy-based proxies.
To ensure that the methodology of our analysis is valid, we perform several robustness checks (see Appendix~\ref{sec:thresholding_analysis}). 

A complementary cross-correlation analysis confirms the relative ordering of the separation distances and provides an independent validation of the peak-matching approach.% (see Appendix \ref{sec:cross-correlation}). 
In addition, it quantifies how the spatial relationships depend on the detection threshold, thereby assessing the robustness of the identified coupling between LET and the kinetic proxies.

These results indicate that peaks of all kinetic dissipation proxies are preferentially located in the vicinity of LET peaks, with distinct characteristic distances depending on the nature of the proxy considered.

\begin{figure*}[ht!]
\includegraphics[width=\textwidth]{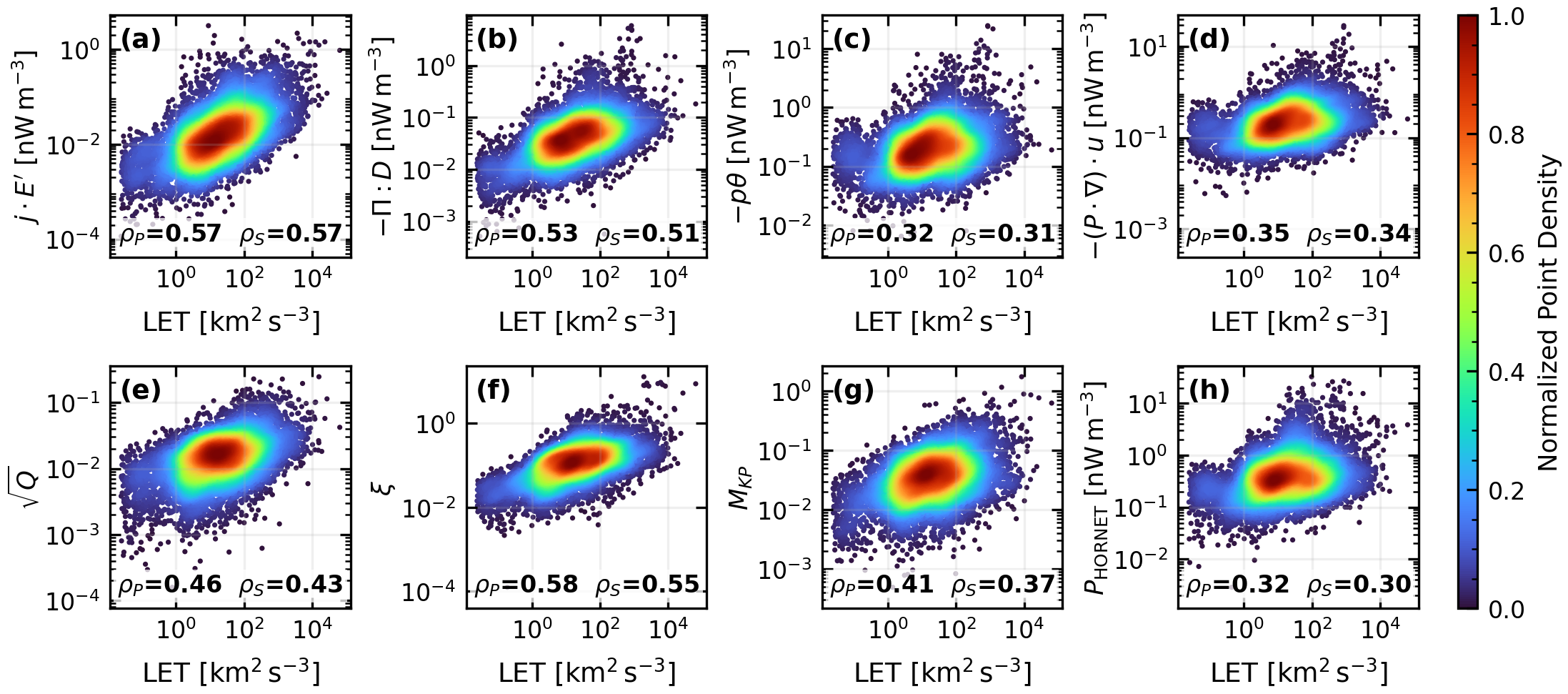}
\caption{Joint probability density functions between LET and kinetic proxies. Pearson ($\rho_P$) and Spearman ($\rho_S$) correlation coefficients are computed. The color bar shows the probability density estimated via kernel density estimation.
\label{fig:jpdf_mean_detrending}}
\end{figure*}

To further investigate the relationship between turbulent activity and kinetic signatures, we examine the correlations between the amplitudes of LET and the corresponding matched kinetic events. 
For each pair of matched peaks, the amplitudes of LET and the dissipation proxy are compared after mean detrending and taking absolute values.
Figure \ref{fig:jpdf_mean_detrending} shows the joint distribution of LET and each dissipation proxy. Each panel reports the Pearson and Spearman correlation coefficients. Pearson measures linear relationships, while Spearman captures monotonic but potentially nonlinear relationships.
Strong correlations are observed between LET and electromagnetic work $\mathbf{J}\cdot \mathbf{E'}$, the shear-driven term $\mathrm{\Pi}-\mathrm{D}$, and non-Maxwellianity $\xi$. 
This is followed by moderate correlations for agyrotropy $\sqrt{Q}$ and entropy deficit $M_{KP}$. Finally, weaker correlations are obtained for the compressive term $P-\theta$ and the $P_{\mathrm{HORNET}}$ proxy.

We also tested the correlation strengths of different detection threshold choices and found that their strengths remain relatively stable with the exception of VDF-based proxies (see Appendix \ref{sec:thresholding_analysis}). 
Additional tests performed across different plasma $\beta$ intervals show that the correlations tend to increase with $\beta$, while remaining well defined within the most populated range of typical magnetosheath conditions (see Appendix \ref{sec:beta_binning}).

Overall, the statistical analysis reveals a systematic spatial clustering of kinetic proxy peaks around LET peaks with characteristic separations at ion scales, and a range of correlation strengths between LET and the different kinetic diagnostics.

\section{Discussion and Conclusions}

The results presented in this study provide new statistical constraints on how turbulent energy is coupled to kinetic processes in the terrestrial magnetosheath. 
Combining an unbiased dataset with multiple complementary diagnostics, we identify a coherent spatial organization of kinetic activity around intermittent turbulent structures at ion scales.

A central result is the emergence of a scale-dependent ordering between different classes of diagnostics. Energy-based proxies show the closest spatial correspondence with LET, with small separations showing that energy conversion occurs in the vicinity of intermittent structures. This supports a picture in which electromagnetic work and pressure–strain interactions act as immediate channels for converting bulk and field energy into particle internal energy at ion scales.

In contrast, velocity-space diagnostics exhibit broader spatial distributions which extend beyond the regions of strongest energy conversion. 
This indicates that the particle response is not confined to these localized sites but instead spreads over larger spatial and velocity-space scales. The additional threshold analysis presented in Appendix~\ref{sec:thresholding_analysis} further indicates that the strongest kinetic signatures are more closely associated with intense intermittent events. Such behavior is consistent with kinetic processes that redistribute energy in phase space, including phase mixing, finite Larmor radius effects, and wave–particle interactions.

Taken together, these observations support a multi-stage scenario. Turbulent interactions first concentrate energy at intermittent structures, energy is then converted locally into particle internal energy, and the resulting particle distributions evolve over larger scales through kinetic processes. This separation between energy transfer, conversion, and particle response is reflected in the finite spatial offsets observed between the corresponding diagnostics.

The coexistence of strong but imperfect correlations further suggests that no single diagnostic fully captures the complexity of kinetic dissipation. Some proxies may include contributions from reversible energy exchanges or depend sensitively on local plasma conditions \citep{Hellinger_2024}. Differences between these proxies likely also reflect their sensitivity to distinct physical processes.

Several limitations should be noted. First, the peak-based approach simplifies the spatial structure of turbulence. Second, the LET is derived under assumptions of isotropy and incompressibility, which are only approximately satisfied in the magnetosheath \citep{pecora2026impact_anisotropy}. Finally, the analysis is statistical and does not establish causality between turbulent structures and kinetic processes.

Despite these limitations, the present results provide robust observational evidence for a structured coupling between turbulence and kinetic processes. Overall, they support a picture in which turbulent energy transfer, kinetic activity, and particle energization are tightly coupled but spatially offset processes. Future work should aim to distinguish between different channels of energy conversion and to separate reversible from irreversible contributions using sign-resolved proxies.

\begin{acknowledgments}
{LSV} was supported by the Swedish National Space Agency (SNSA) Research Grant N. 2025-00229.
{LSV and OP} were supported by the Swedish Research Council (VR) Research Grant N. 2022-03352 and by the PRIN 2022 project ``2022KL38BK - The ULtimate fate of TuRbulence from space to laboratory plAsmas (ULTRA)'' (Master CUP B53D23004850006) by the Italian Ministry of University and Research, funded under the National Recovery and Resilience Plan (NRRP), Mission 4 – Component C2 – Investment 1.1, ``Fondo per il Programma Nazionale di Ricerca e Progetti di Rilevante Interesse Nazionale (PRIN 2022)'' (PE9) by the European Union – NextGenerationEU. 
{OP} is supported by the FIS2 Starting Grant FIS-2023-00246 “PhAse-sPAce cOmplexity in turbulent nearly-reversible plasmas (PAPAO)” (CUP B53C24009610001) funded by the Italian Ministry of University and Research.
{AT} was supported by the ...  {ZV} was funded by the Austrian Science Fund (FWF) under 10.55776/PAT9232923. {FP} is partially supported by NSF SHINE 2501387 at the University of Delaware. TV is supported by NASA-FINESST grant No. 80NSSC23K1630. {SS} is supported by ``Space It Up!'', funded by the Italian Space Agency (ASI) and the Ministry of University and Research (MUR), under Contract Grant Nos. 2024-5-E.0-CUP and I53D24000060005.
{VMC} is supported by the Vicen\c{c} Mut grant POSTDOC2024\_58 funded by Conselleria d’Educació i Universitats del Govern de les Illes Balears i pel Fons Social Europeu +.

\end{acknowledgments}

\begin{contribution}
%%This section gives authors the space to recognize author contributions. The text inside this environment is NOT counted towards the total word quanta. At a minimum, manuscripts are expected to include this text:

LVAD, LS and PR contributed equally to all the elements of the work. VMC and AS contributed to the development of data analysis codes and to the interpretation of results. All other authors contributed to the data analysis, interpretation of results, and manuscript preparation.

%% But authors are expected to provide more specific details, e.g. 
%%
%%SC was responsible for writing and submitting the manuscript.
%%WWM came up with the initial research concept and edited the manuscript.
%%OTS obtained the funding and edited the manuscript.
%%EBF provided the formal analysis and validation. He also edited the manuscript.
%%GEH Supervised the undergraduates, wrote the software and administers the project github and Zenodo repositories.
%%
%% Authors can use the Contributor Role Taxonomy (CRediT) at
%% for ideas on how write a good statement tailored to their needs.

\end{contribution}


%%%%\facilities{HST(STIS), Swift(XRT and UVOT), AAVSO, CTIO:1.3m, CTIO:1.5m, CXO}

\software{Pyrfu \citep{pyrfu_2024}}

\appendix
\shorttitle{Correlations between turbulent structures and dissipation in the magnetosheath}
\shortauthors{Doan et al.}

\section{Residual spintone filtering}
\label{sec:spintone}

\begin{figure*}[ht!]
\includegraphics[width=\textwidth]{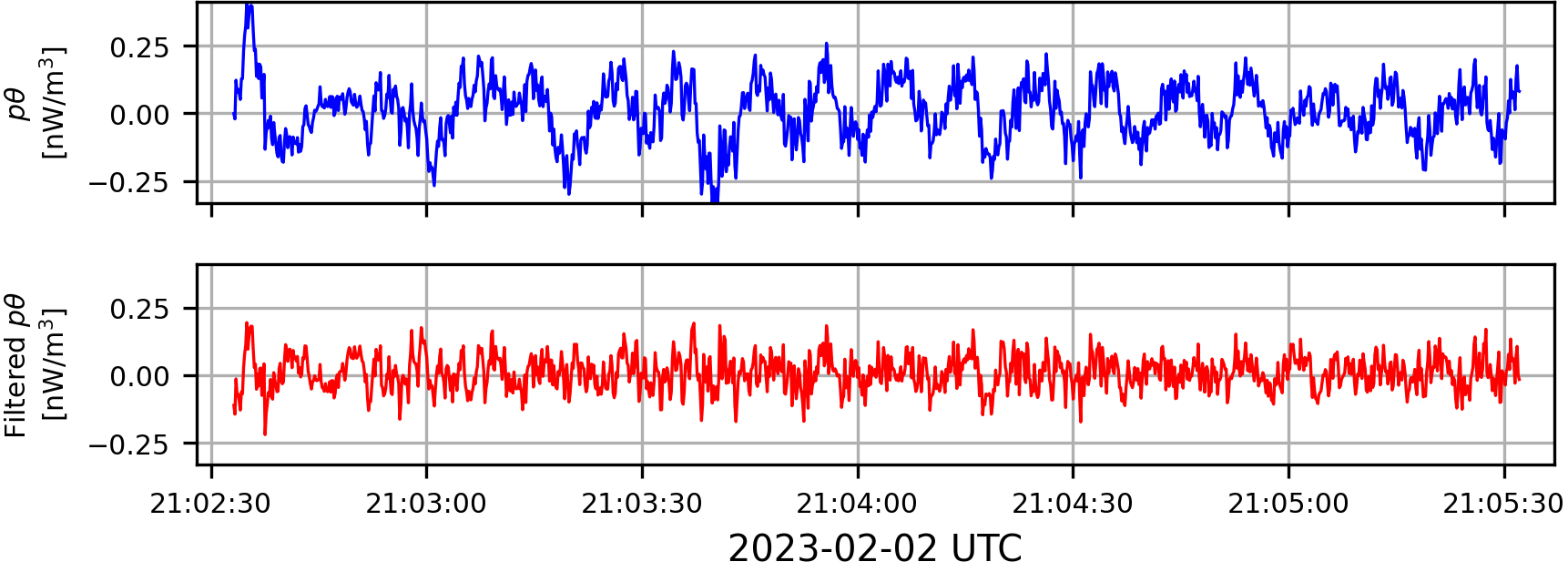}
\caption{
High-pass filtering applied to the $p\theta$ proxy. After applying the filter with a cutoff frequency of $1.05 \times f_{\text{dominant}}$ to the original signal (blue), the filtered signal (red) effectively suppresses these oscillations while preserving the physical trends.}
\label{fig:filtering}
\end{figure*}

During the computation of gradient-related quantities such as $Pi-D$ and $p\theta$, significant oscillations were observed and are attributed to residual spintone effects (see Figure~\ref{fig:filtering}). These oscillations exhibit regular patterns, and their dominant frequency was identified using a Welch power spectral density estimate. To mitigate their impact, a high-pass Fourier filter was applied and targeted frequencies above the dominant frequency with a $10\%$ margin. This margin accounts for numerical precision issues, as filtering at the exact dominant frequency sometimes fails to suppress the oscillations effectively.

\section{Methodology for the spatial displacement matching between peaks} 
\label{sec:method_distance}

Peak detection is performed using a threshold defined relative to the local statistical variability of the signal. For each diagnostic time series $x(t)$, the mean $\mu$ and standard deviation $\sigma$ are computed over the considered interval. A detection threshold $n$ is then applied to identify peak times $t_{\mathrm{peak}}$ satisfying
$$
|x(t_{\mathrm{peak}}) - \mu| > n\sigma .
$$
Only peaks exceeding this threshold are considered candidate events. To prevent clusters of nearby local maxima from being counted as independent structures, we additionally apply prominence filtering from Python's scipy library.

To quantify spatial separations between turbulent structures and kinetic signatures, we match peaks detected in the LET signal with peaks detected in each kinetic proxy. Since multiple peaks may occur within a short interval, a one-to-one matching procedure is required to avoid ambiguous associations. Let the detected peak times of signals $A$ and $B$ be respectively
$P_A = \{ t_{A,i} \}_{i=1}^{n}$ and $P_B = \{ t_{B,j} \}_{j=1}^{m},$ with $t_{A,i}, t_{B,j} \in \mathbb{R}$. We construct a one-to-one matching using a mutual nearest-neighbor criterion. The algorithm proceeds as follows:

\begin{enumerate}
\item \textbf{Nearest neighbor from $A$ to $B$}: For each peak $t_{A,i} \in P_A$, define its nearest neighbor in $P_B$ by \[
\pi(i) = \arg\min_{1 \le j \le m} 
\left| t_{A,i} - t_{B,j} \right|.
\]
This yields candidate pairs $(t_{A,i}, t_{B,\pi(i)})$. At this stage, multiple peaks in A may map to the same peak in B. We note that in the case of a tie-breaker distance, we choose the peak $t_{B,\pi(i)}$ with the largest amplitude. if still tied, choose the smallest index. 

\item \textbf{Nearest neighbor from $B$ to $A$}: Similarly, for each peak $t_{B,j} \in P_B$, define its nearest neighbor in $P_A$ by
\[
\sigma(j) = \arg\min_{1 \le i \le n} 
\left| t_{A,i} - t_{B,j} \right|.
\]
The same tie-breaker rule is applied if necessary.

\item \textbf{Mutual matching}: A pair $(t_{A,i}, t_{B,j})$ is retained if and only if
\[
\pi(i) = j 
\quad \text{and} \quad 
\sigma(j) = i.
\]
\end{enumerate}

The final matched set is
\[
M 
= 
\left\{
(t_{A,i}, t_{B,j}) \in P_A \times P_B
\;:\;
\pi(i) = j 
\text{ and }
\sigma(j) = i
\right\}.
\]

This process ensures a one-to-one correspondence between peaks and prevents spurious matches caused by asymmetric or distant assignments. All peaks that do not satisfy the mutual matching step are discarded from the analysis.

\begin{figure*}[ht!]
\includegraphics[width=\textwidth]{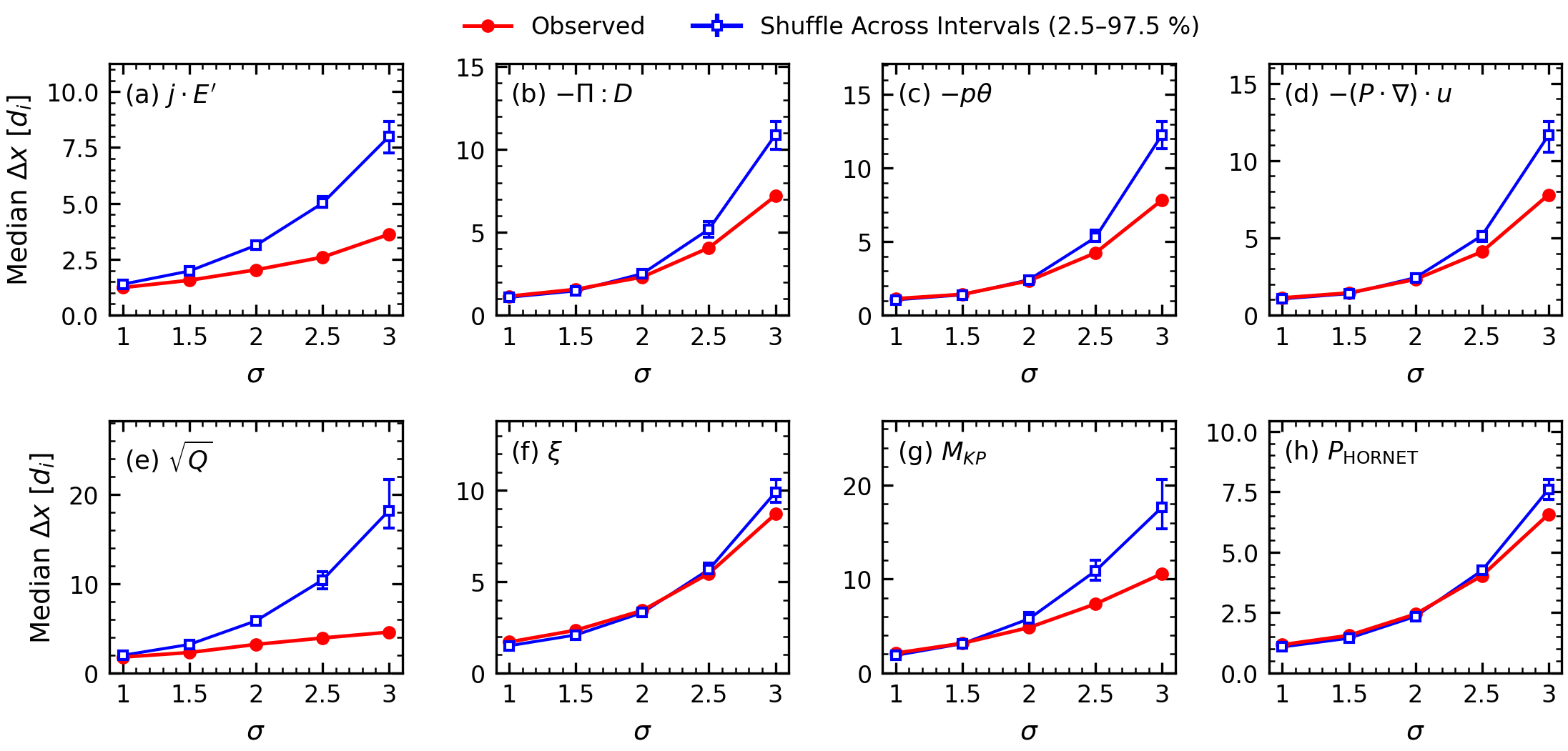}
\caption{
Left: Mean spatial peak distance between LET and kinetic proxies as a function of threshold for the unbiased campaign dataset. Right: Same analysis but by shuffling the dataset of each kinetic proxy within each interval and preserving LET.
}
\label{fig:distance_compare}
\end{figure*}

To assess the significance of the observed spatial associations between LET peaks and kinetic proxy peaks, we perform a null hypothesis test by randomly shuffling the kinetic proxy signals and repeating the peak detection and matching procedure (see Figure~\ref{fig:distance_compare}). At low thresholds, both the shuffled dataset (right) and the real data (left) show similar trends due to the high density of available peaks. In contrast, at higher thresholds, the shuffled dataset systematically yields larger peak distances and a wider spread than the real data. This demonstrates that the closer spatial alignment observed in the real dataset is not reproduced by random shuffling. We note that correlations persist at all thresholds in the shuffled data. This is expected as the thresholding step ensures similar peak amplitudes are detected regardless of signal shuffling.

\section{Influence of the thresholding on the detected structures} 
\label{sec:thresholding_analysis}

The detection threshold $n$, expressed in units of the local standard deviation $\sigma$, controls the amplitude of the selected events and therefore probes different levels of intermittency. Lower thresholds include weaker fluctuations, while higher thresholds isolate the most intense structures.

The left panel of Figure 6 shows that the correlations between LET and the dissipation proxies remain largely stable across the explored range of thresholds. This indicates that the coupling between turbulent energy transfer and kinetic activity is not dominated by rare extreme events, but instead reflects a persistent property of the turbulent fluctuations. The relative ordering between proxies is preserved and demonstrates that the hierarchy identified in Section \ref{sec:statistics} is robust to event selection.

A distinct behavior is observed for the velocity-space diagnostics $\xi$ and $M_{KP}$, whose correlations with LET increase with the detection threshold. This indicates that strong deviations from equilibrium are preferentially associated with the most intense turbulent structures.

\begin{figure*}[ht!]
\includegraphics[width=\textwidth]{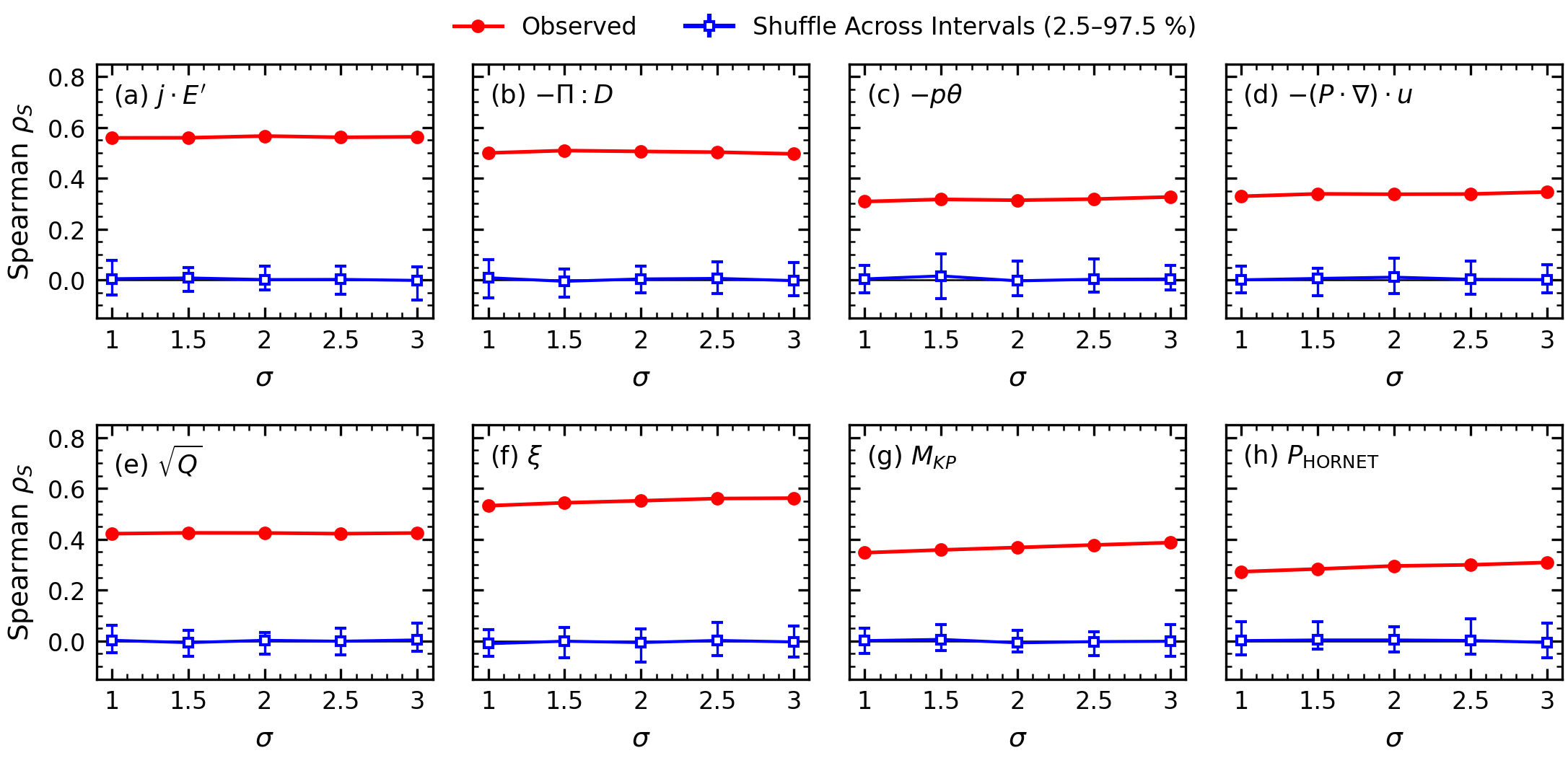}
\caption{
Left: Maximum correlation between LET and the each kinetic proxy as a function of the detection threshold expressed in units of the standard deviation $\sigma$. Right: Maximum correlation for $\xi$ and $M_{KP}$ using different velocity-space integration during their computation.}
\label{fig:corr_thres}
\end{figure*}

To assess whether this trend could be influenced by measurement noise in the VDFs, we restrict the velocity-space integration to $0.5v_{th} < v < 6.8v_{th}$. As shown in the right panel of Figure~\ref{fig:corr_thres}, this restriction systematically reduces the correlations with LET. If the excluded regions were dominated by counting noise, then their removal would reduce variance and increase correlations. The observed opposite behavior therefore indicates that these regions contain physically meaningful distortions associated with turbulent structures. This supports the interpretation that the enhanced correlations at higher thresholds reflect genuine kinetic effects rather than instrumental artifacts, and justifies the use of the full VDF in the main analysis.

\begin{figure*}[ht!]
\includegraphics[width=\textwidth]{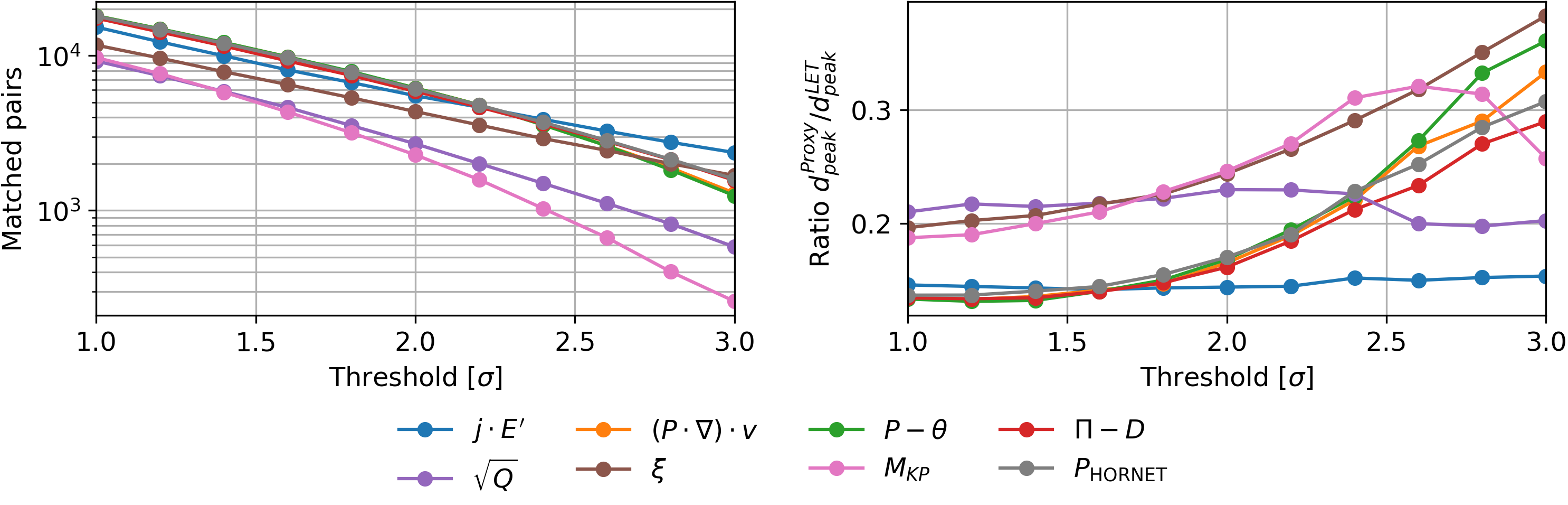}
\caption{
Influence of the detection threshold on the pairing statistics between LET and the kinetic proxies. 
The left panel shows the number of matched structures as a function of the detection threshold, while the right panel shows the ratio between the median separation distance of proxy events $d_{peak}^{Proxy}$ and the median distance of successive LET events $d_{peak}^{LET}$. }
\label{fig:pair_thres}
\end{figure*}

Finally, Figure \ref{fig:pair_thres} shows that increasing the threshold reduces the number of detected events but does not significantly modify their spatial correspondence with LET structures. The persistence of clustering across thresholds indicates that the localization of kinetic activity near turbulent structures is not an artifact of event selection. Based on this analysis, 
$n=2$ provides a compromise between statistical convergence and the identification of sufficiently intermittent events.

\section{Turbulent-kinetic correlations across plasma $\beta$ regimes}
\label{sec:beta_binning}

To assess the dependence of the results on plasma conditions, we repeat the analysis across different ranges of the plasma $\beta$, which is known to influence the nature of turbulence and kinetic processes in the magnetosheath.

\begin{figure*}[ht!]
\includegraphics[width=\textwidth]{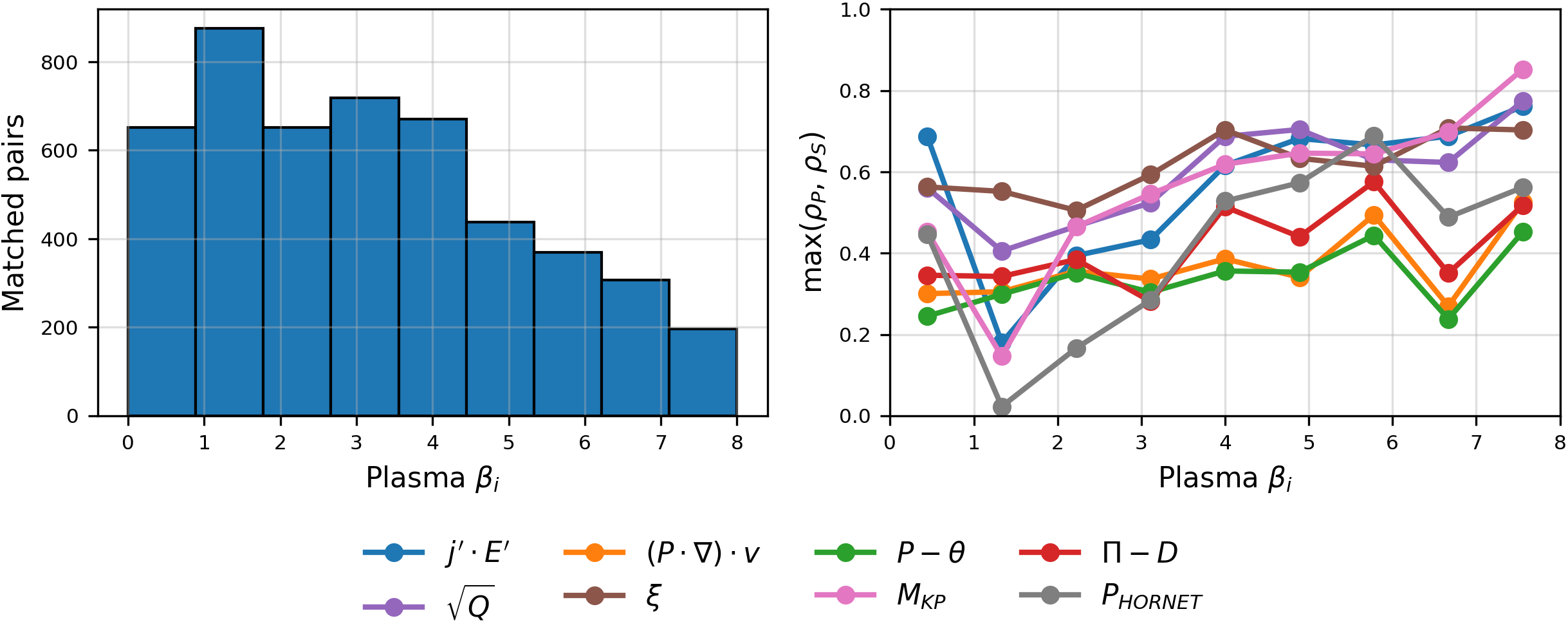}
\caption{
Turbulent–kinetic correlations across plasma $\beta$ regimes. The left panel shows the probability density function of intervals as a function of $\beta$, normalized by the total number of intervals. The right panel shows the corresponding maximum correlations (Pearson or Spearman) between LET and each kinetic diagnostics as a function of $\beta$. }
\label{fig:beta_dependence}
\end{figure*}

Figure~9 shows that the correlations between LET and the kinetic proxies tend to increase with plasma $\beta$. The binning is constructed to account for the bin width to ensure that the reported values correspond to properly normalized probability densities and allowing for a consistent comparison across $\beta$ regimes.

The observed increase in correlation at higher $\beta$ indicates a stronger coupling between turbulent fluctuations and kinetic processes in these regimes. High-$\beta$ plasmas are commonly associated with enhanced compressibility and stronger pressure fluctuations. Consequently, it can promote more efficient energy conversion through pressure-strain interactions and facilitate the development of velocity-space distortions. This behavior is consistent with previous studies showing that the partition of energy conversion channels depends on plasma $\beta$ \citep{Parashar_2018}.

At the same time, non-negligible correlations are observed across all $\beta$ intervals and indicate that the coupling between turbulent structures and kinetic activity is not restricted to specific plasma conditions. Although the dataset is predominantly composed of intervals in the range $1 \lesssim \beta \lesssim 2$, the persistence of correlations outside this regime demonstrates that the observed trends are not driven by a particular subset of the data.

Overall, these results indicate that the coupling between turbulence and kinetic processes is a robust feature of magnetosheath plasma, with an efficiency that increases with $\beta$.

\bibliography{sample701}{}
\bibliographystyle{aasjournalv7}



\end{document}